# Nonzero angular momentum pairing correlation in shell model

S. Haq , Y. Sadeq and I. M. Hamammu[1]

Department of Physics, Faculty of Science, P.O. Box 9480, University of Garyounis,
Benghazi, Libya, Fax 00218612240757
[1]Email:Hamammu@Garyounis.edu

## Abstract

A simple approximation to shell model is proposed in which the low energy excitation spectra corresponds to the identical nucleons occupying the same single particle states where they preferred to form pairs for the ground states. We call this approximation as nonzero angular momentum pairing shell model. It not only reduces the dimensionality of the shell model but also matches the number of low energy levels in experimental spectra for few cases where exact shell model predicts many more states. The special focus has been done to consider the realistic interaction derived from free nucleon-nucleon scattering data to cope with the experimental spectra. The proposed approximation to shell model has been applied to calculate the energy spectra of $O^{18}$ and $Ni^{58}$ nuclei where only two neutrons occupy the valence states outside the core. When compared with the experimental data, the results are found to be encouraging. It is expected that results will be more pronounced if the even-even nuclei with higher number of valence nucleons are considered.

## I. Introduction

It is well known that among all nuclear models, the shell model is more basic and realistic as it describes the nucleus in terms of its stable constituents – the neutrons and the protons. However all is not well with the shell model. The reasons are as follows.

(1) By increasing number of valence states and valence particles the dimensionality of the Hamiltonian matrices increases beyond limits. As an example we show here the case of $Sm^{154}$ nucleus cited by Talmi [1]. Considering for the nucleus $_{62}Sm^{154}_{92}$ , $_{50}Sn^{132}_{82}$ as core, one finds 12 protons and 10 neutrons distributed among valences shells. For protons the states are $0g_{7/2}$ , $1d_{5/2}$, $2S_{1/2}$, $1d_{3/2}$, $0h_{11/2}$ and for neutrons states are $0g_{7/2}$ , $0h_{11/2}$ , $1f_{7/2}$ , $2p_{3/2}$, $1f_{5/2}$, $2p_{1/2}$, $0i_{13/2}$ respectively. The various dimensions of the Hamiltonian matrices for low positive parity states are shown in Table(1)

**Table(1)** Column 1 shows angular momentum with parity, and column 2 gives the corresponding dimensions of Hamiltonian matrices.

| J | Number of dimensionality |
|---|---|
| $0^+$ | 41,654,193,516,797 |
| $2^+$ | 346,132,052,934,889 |
| $4^+$ | 530,897,397,260,575 |

For J=$0^+$, the order of the Hamiltonian matrix to be diagonalze is $10^{14}$ .To construct such matrix needs 1200 diagonal and non-diagonal effective interaction matrix elements. It is impossible to diagonalize such matrices with modern computers easily. In fact there may be very little use of $10^{14}$ components of the wave function for the nucleus concerned. Now a days large scale shell model calculations are being performed [2]. But the question still remains: "Are all these components in the wave functions are really important and required?". To overcome these difficulties many approximate methods based on the guidance of strong pairing correlations among nucleons have been proposed and successfully implemented. The examples of such types of approximations to shell model are low seniority shell model [3], broken pair model [4], [5] and nucleon pair shell model [6] etc. In this work we give a new proposal that is to restrict the valence particles pair-wise in the same state. Next section is devoted for the same.

(2) There is no prefect guidance for the effective matrix elements required in the shell model calculations. This is due to the fact that unlike well known Coulomb and gravitational interactions, the exact nature of nuclear interaction is not known. Although one knows the various properties of nuclear interactions such as:

    (i)       It is short range ~ $2 \times 10^{-13}$ cm which is 2 Fermi (F).

    (ii)      It is spin dependent i.e. different for spin singlet and triplet states.

    (iii)     It is charge independent i.e. it is same for neutron-neutron, proton-proton and neutron- proton.

    (iv)     It contains a tensor part in it i.e. it not only depends on radial distance r but also depend on angle $\theta$ and $\phi$.

    (v)      It has hard core i.e. repulsive at very short range ~ 0.4F.

    (vi)    It may depend on velocity.

        But unfortunately even when these aspects are incorporated in the derivation of realistic set of matrix elements, the result of energy spectra comes out to be poorer

compare to the other set of parameterized matrix elements such as empirical and modified surface delta interaction.

The nonzero angular momentum pairing correlation approximation has been applied to calculate low energy spectrum of two nuclei i.e. the $O^{18}$ and $Ni^{58}$ where only two neutrons occupy the valence states outside the appropriate cores. For the effective two body interaction we have chosen the well known Tabakin potential (TP) [7] as appropriately calculated in the form of effective interaction by Kuo et al [8]. The Tabakin potential is based on free nucleon-nucleon scattering data up to 320 MeV and is parameterized to reproduce the scattering phase shifts. The result is encouraging when compared with the low energy experimental spectra.

## II. Nonzero pairing correlations

In nuclear physics a nucleon pair is defined when two protons or two neutrons occupy the same state j but opposite projection quantum numbers m i.e. if one has state $\left| j\ m \right\rangle$ then another has state $\left| j\ -m \right\rangle$. When two nucleons are correlated like this then their total angular momentum J obviously becomes zero. When a pair is broken then they can occupy any state of the type $\left| j_1\ m_1 \right\rangle$ and $\left| j_2\ m_2 \right\rangle$ without any restrictions. We propose here that for low energy excitations they still occupy the same state j but different m. These components already exist in exact shell model calculations. We only discard those components which belong to the different states. The justification for our proposal lies in the fact that:

(1) At low energy for many nuclei experimentally there are only few excited states which exist whereas exact shell model gives many undesired states especially when the model space is large.

(2) Realistic matrix elements are usually evaluated in L-S coupling scheme. For shell model calculation we have to transfer them to j-j coupling scheme. For the particle occupying the same state j, the state is shown below.

$$\left|(j_1)^2:JT\right\rangle=\frac{1}{2}\sum_{LS}\sum_{N_{cm}L_{cm},n\ell,\Im}\begin{bmatrix}\ell_1 & \frac{1}{2} & j_1\\ \ell_1 & \frac{1}{2} & j_1\\ L & S & J\end{bmatrix}\left\langle N_{cm}L_{cm},n\ell:L\|n_1\ell_1,n_1\ell_1:L\right\rangle$$

$$\times\left[1-(-1)^{\ell+S+T}\right]U(L_{cm}\ell\ J\ S:L\ \Im)\left|N_{cm}L_{cm},(n\ell,S)\Im:JT\right\rangle$$

$$\ldots\ldots\ldots\ldots\ldots\ldots, \quad (1)$$

where various entities are defined in Ref.[9]. Decomposing above as

$$\left|(j_1)^2:JT\right\rangle=\left|(j_1)^2:JT\right\rangle_{S=0}+\left|(j_1)^2:JT\right\rangle_{S=1},$$

where each term takes the form as follows

$$\left|(j_1)^2:J\ T\right\rangle_{S=0}=\frac{1}{2}\sum_{L}\sum_{N_{cm}L_{cm},n\ell,\Im=\ell}\begin{bmatrix}\ell_1 & \frac{1}{2} & j_1\\ \ell_1 & \frac{1}{2} & j_1\\ L & 0 & L\end{bmatrix}\left\langle N_{cm}L_{cm},n\ell:L\|n_1\ell_1,n_1\ell_1:L\right\rangle$$

$$\times\left[1-(-1)^{\ell+T}\right]U(L_{cm}\ell\ L\ 0:L\ \ell)\left|N_{cm}L_{cm},(n\ell,0)\ell:L\ T\right\rangle$$

$$(2.a)$$

$$\left|(j_1)^2:J\ T\right\rangle_{S=1}=\frac{1}{2}\sum_{L}\sum_{N_{cm}L_{cm},n\ell,\Im}\begin{bmatrix}\ell_1 & \frac{1}{2} & j_1\\ \ell_1 & \frac{1}{2} & j_1\\ L & 1 & J\end{bmatrix}\left\langle N_{cm}L_{cm},n\ell:L\|n_1\ell_1,n_1\ell_1:L\right\rangle$$

$$\times\left[1+(-1)^{\ell+T}\right]U(L_{cm}\ell\ L\ 1:L\ \Im)\left|N_{cm}L_{cm},(n\ell,1)\Im:J\ T\right\rangle \quad .$$

$$(2.b)$$

The overlap of these wave functions for the states J $= 0^+$ when T=1 are shown in Table (2) and the calculated numerical values are given in Tables (3) and (4). Looking at Tables (3) and (4) we notice that the contribution of L=S=1 to J $=0^+$ is large enough for each states. This state can couple to J $=2^+$ also without moving from the same j.

**Table(2)** Overlap for pairing states when J $=0^+$ and T=1 with respect to each L and S .

| J | L | S | $\left\langle (j_1)^2:0,T=1\,\middle|\,(j_1)^2:0,T=1\right\rangle$ | |
|---|---|---|---|---|
| $0^+$ | 0 | 0 | $\dfrac{1}{2}\left(\dfrac{[j_1]}{[\ell_1]}\right)\displaystyle\sum_{N_{cm}\ell,\,n\ell}\left\langle N_{cm}\ell,n\,\ell:0\,\middle\|\,n_1\ell_1,n_1\ell_1:0\right\rangle^2$ | ; $\ell$ even |
| | 1 | 1 | $\left(\dfrac{[j_1]}{2}\right)\dfrac{\left\{\ell_1\left[\frac{1}{2}\ell_1\right]-j_1\left[\frac{1}{2}j_1\right]+\frac{3}{4}\right\}^2}{\ell_1\left[\frac{1}{2}\ell_1\right][\ell_1]}\displaystyle\sum_{N_{cm},\,n\ell}\left\langle N_{cm}\ell,n\ell:1\,\middle\|\,n_1\ell_1,n_1\ell_1:1\right\rangle^2$ | ; $\ell$ odd |

**Table(3)** The numerical overlap for pairing states 1p$_{3/2}$ and 0f$_{5/2}$ when J $=0^+$ and T=1 with respect to each L and S.

| J | L | S | $\left\langle (1p_{1/2})^2:\Omega,T=1\,\middle|\,(1p_{1/2})^2:\Omega,T=1\right\rangle$ | $\left\langle (0g_{9/2})^2:J,T=1\,\middle|\,(0g_{9/2})^2:J,T=1\right\rangle$ |
|---|---|---|---|---|
| $0^+$ | 0 | 0 | 0.3333 | 0.5556 |
| | 1 | 1 | 0.6667 | 0.4444 |

**Table(4)** The numerical overlap for pairing states 1p$_{1/2}$ and 0g$_{9/2}$ when J $=0^+$ and T=1 with respect to each L and S.

| J | L | S | $\left\langle (1p_{3/2})^2:\Omega,T=1\,\middle|\,(1p_{3/2})^2:\Omega,T=1\right\rangle$ | $\left\langle (0f_{5/2})^2:J,T=1\,\middle|\,(0f_{5/2})^2:J,T=1\right\rangle$ |
|---|---|---|---|---|
| $0^+$ | 0 | 0 | 0.6667 | 0.4286 |
| | 1 | 1 | 0.3333 | 0.5714 |

Experimentally, the first excited state of all even-even nuclei is $2^+$ state except for few doubly magic nuclei where the first excited state is $0^+$. This excited $2^+$ state in terms of L-S coupling is clearly L =1 and S=1 state when the pairs are in the same state j. It is to be remarked here that this very experimental fact has been incorporated in highly successful interacting bosons model (IBM).

Owing to the above discussions we propose here especially for low energy spectra to restrict the particles in the same state j.

## III.  Application of Nonzero Pairing Shell Model (NZPSM) to  Ni$^{58}$& O$^{18}$

## III.1 The nucleus Ni$^{58}$

For this nucleus, Ni$^{56}$ has been chosen as core corresponding to magic number Z=28 and N=28 i.e. Ni$^{58}$=Ni$^{56}$+2n. The two neutrons are first distributed in three single particle (s.p.) states $1p_{3/2}$, $0f_{5/2}$ and $1p_{1/2}$ and then four s.p. states $1p_{3/2}$, $0f_{5/2}$, $1p_{1/2}$ and $0g_{9/2}$. We have taken s.p. energies as shown in Table (5) from experimental spectra of Ni$^{57}$ i.e. Ni$^{56}$+1n [10]. The energy spectra using ESM and NZPSM after diagonalzing Hamiltonian matrices [11] along with the experimental (EXPT) data are shown in Fig.(2) for realistic Tabakin potential and in Fig.(3) for MSDI. It is to be noted here that for J =0$^{+}$ ESM and NZPSM results coincide. Looking at Fig.(2) for  TP we find that for J  = 2$^{+}$,1.77 MeV, 2.10 MeV and 2.97 MeV states from ESM are extra configurations not found in experimental spectra as mentioned earlier. These states come due to ( $1p_{3/2}$, $0f_{5/2}$ ),( $0f_{5/2}$, $1p_{1/2}$ ) and ( $1p_{3/2}$, $1p_{1/2}$ ) configurations. Similarly for J =4$^{+}$, 1.71 MeV extra state comes from ($1p_{3/2}$, $1p_{1/2}$) configuration. Moreover the first two, 2$^{+}$ excited states have been lifted up comparing little better with EXPT as compare to ESM.

Similar conclusion can be drawn for MSDI for missing 2$^{+}$ state at 2.59 but NZPSM result is not as good as compare to TP. The reason is obvious as MSDI is a set of m.e. which is obtained by fitting the strength parameters which depend on neglected configurations, whereas realistic interaction set of TBME is a fit from nucleon-nucleon scattering data.

**Table (5)** The single particle energies in MeV of Ni$^{57}$ .

| States | $1p_{3/2}$ | $0f_{5/2}$ | $1p_{1/2}$ | $0g_{9/2}$ |
|--------|------------|------------|------------|------------|
| E (MeV) | 0 | 0.78 | 1.08 | 3.5 |

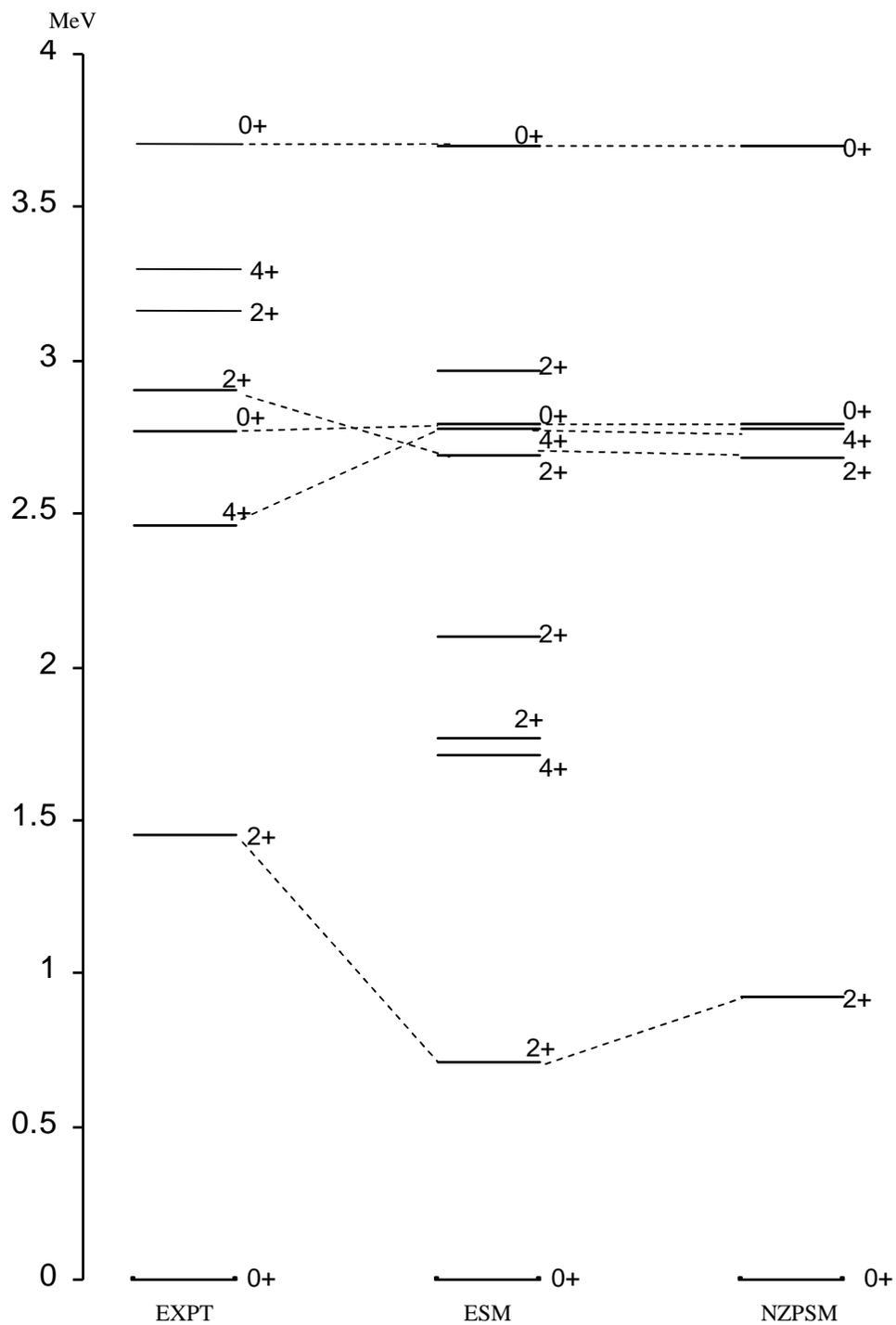

**Fig.(2)** Experimental and calculated energy levels of Ni$^{58}$ using TP, with three valence states 1p$_{3/2}$, 0f$_{5/2}$ and 1p$_{1/2}$. Here EXPT, ESM and NZPSM denote the experimental, exact shell model and nonzero pairing shell model results respectively.

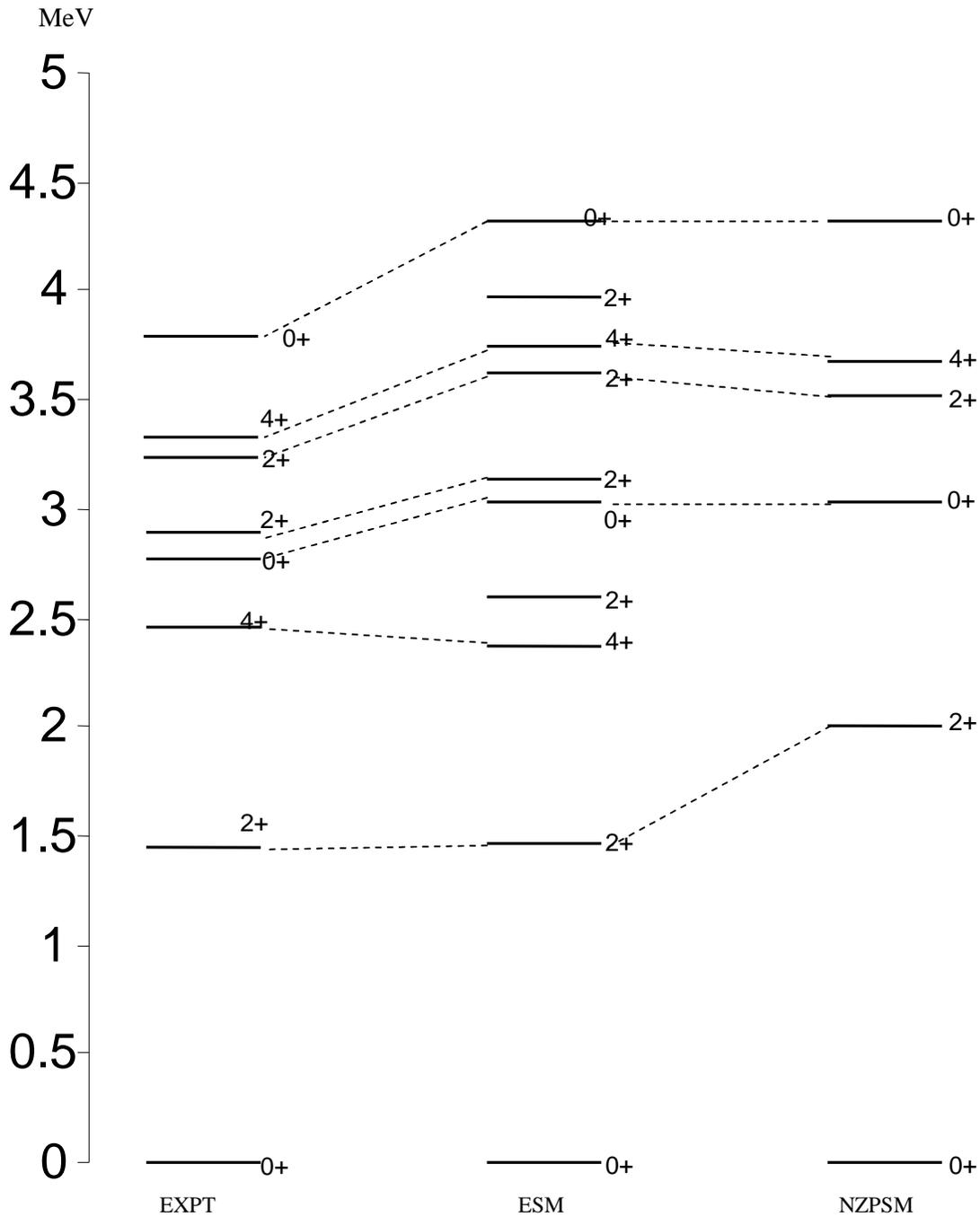

**Fig.(3)** Experimental and calculated energy levels of Ni$^{58}$ using MSDI, with three valence states $1p_{3/2}$, $0f_{5/2}$ and $1p_{1/2}$. Here EXPT, ESM and NZPSM denote the experimental, exact shell model and nonzero pairing shell model results respectively.

By including $0g_{9/2}$ state also in valence space nothing new happens for positive parity states as no additional configurations results in ESM as compare to NZPSM. The resulting energy levels with TP are shown in Fig.(4). Looking at Fig.(4) we notice that including higher s.p. states there are improvement in first and second excited $2^+$ states as little bit they have been lifted up in both ESM and NZPSM. But the lift is more in case of NZPSM for first $2^+$ state as compare to ESM in the sense that they are more close to experimental values. Similar conclusion is drawn for $4^+$ state. However lowest $4^+$ state in ESM is due to ($1p_{3/2}$, $0f_{5/2}$) configuration not present in NZPSM. More over two extra $2^+$ states with energies 1.89 MeV and 2.23 MeV correspond to higher configurations not included in NZPSM.

## III.2 The nucleus $O^{18}$

For this nucleus we have chosen $O^{16}$ as core with respect to Z=8 and N=8 magic numbers and then two neutrons of $O^{18}$ have been distributed in valence s.p. states $0d_{5/2}$ $1s_{1/2}$ and $0d_{3/2}$. The s.p. energies of these states are taken from the experimental spectra of $O^{17}$ [9] i.e. $O^{16}$+1n and are shown in Table (6). For effective two body m.e. again we have chosen TP. The various resulting configurations of wave functions for J =$0^+$, $2^+$ and $4^+$ in ESM and NZPSM are same as for three s.p. states of $Ni^{58}$. Only difference lies in principal (n) and orbital angular momentum ($\ell$) quantum numbers. The final J is same.

The Hamiltonian matrices have been set up and diagonalized for J =$0^+$, $2^+$ and $4^+$ states. The results are given in Fig.(5). Experimental energy spectra of $O^{18}$ is bit complicated as seen from figure. If we see up to 4 MeV then it resembles the vibrational spectra of phonons as $0^+$, $2^+$ and then $0^+$, $2^+$ and $4^+$ states, though $4^+$ is lower in energy. If we go further up then we see another band of $0^+$, $2^+$ and $4^+$. Looking at ESM we notice that up to 4 MeV all states are in accordance with experimental states but lowers in energies. As remarked earlier, usually the results with realistic interactions are not as good as compare to other effective interactions such as empirical and schematic interactions. Looking at NZPSM we find that first three excited states $2^+$, $4^+$ and $0^+$ are exactly in order with experimental states, and are lifted up towards experimental values i.e. little better than ESM. It seems that ESM $2^+$ state at 2.08 MeV (EXPT 3.93 MeV) is a state either coming from ($0d_{5/2}$, $0d_{3/2}$) or ($0d_{3/2}$, $1s_{1/2}$) or ($0d_{5/2}$, $1s_{1/2}$). Naturally it seems that there is another mechanism involved in the excitation process of $O^{18}$ nucleus.

**Table (6)** The single particle energies in MeV of $O^{17}$.

| States | $0d_{5/2}$ | $1s_{1/2}$ | $0d_{3/2}$ |
|--------|--------|--------|--------|
| E (MeV) | 0 | 0.87 | 5.08 |

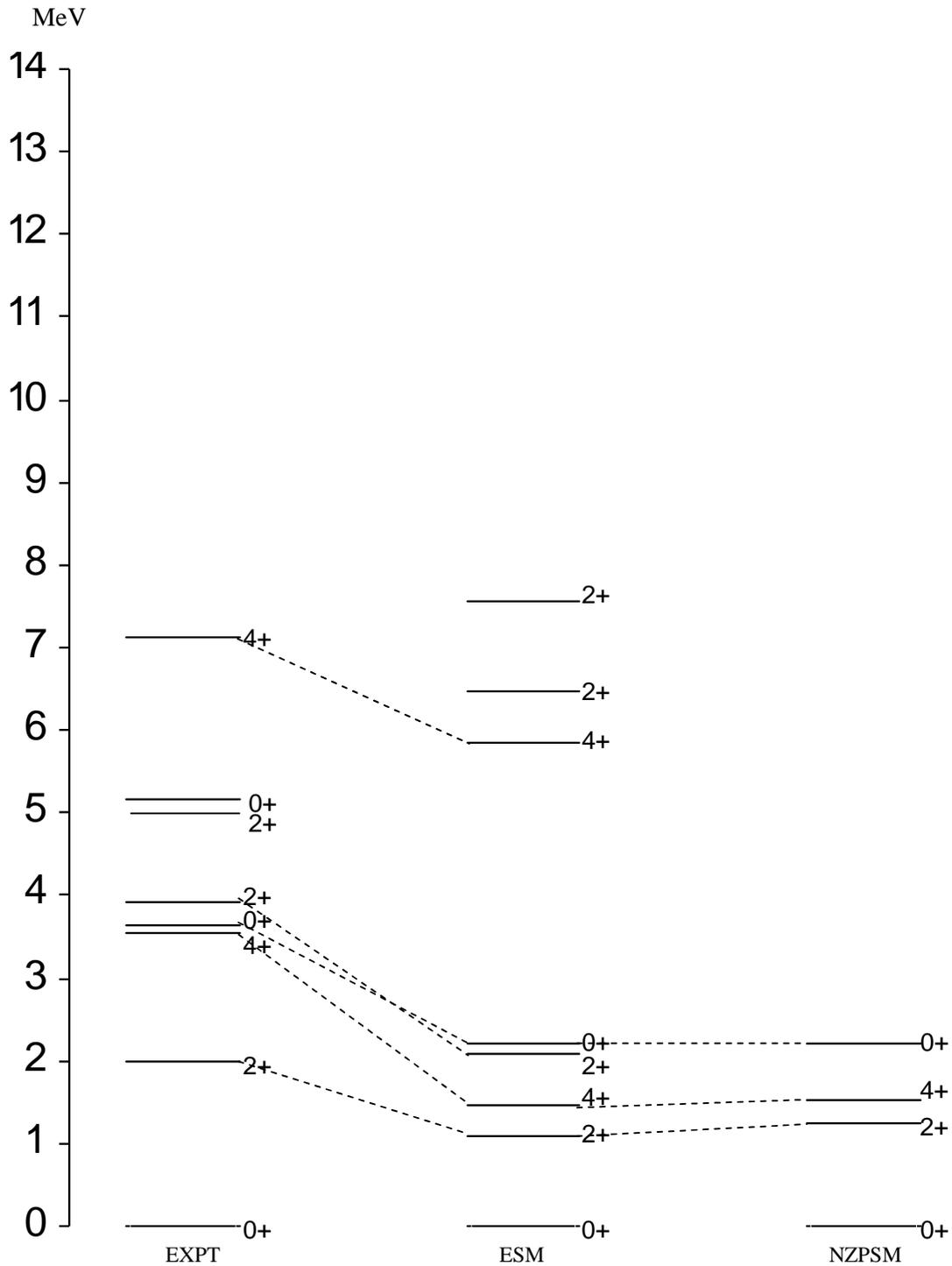

**Fig.(5)** Experimental and calculated energy levels of ESM and NZPSM for $O^{18}$ using TP, with 3 valence states $0d_{5/2}$, $1s_{1/2}$ and $0d_{3/2}$.

## IV.    Conclusion

As mentioned earlier that exact shell model calculations need the choice of proper valence space, s.p. energies of the states involved in valence space and finally the appropriate effective two body interactions. As regards to the effective interaction naturally it should be realistic in nature which is derived from free nucleon-nucleon scattering data. But when these types of interactions are used in ESM, the results are usually unsatisfactory. It does not mean that fault lies in realistic interactions, but rather it lies in ESM methodology where many unnecessary components in wave functions produces extra  states and also lowers the low energy states. Out of many possibilities of nucleon correlations, we presented here a simple one i.e. to restrict the particles in the same j states. By doing this we have found that:

(1) Some of the low energy states which are not present in experimental spectra simply disappear from ESM.

(2) ESM low energy states can be lifted up making them more close to experimental values using realistic interactions.

It is hoped that conclusion found above will be more pronounced once we go for the NZPSM calculation using higher number of particles in the valence shell i.e. 4,6,8…etc. for even nuclei.